\documentclass[twocolumn,showpacs,preprintnumbers,amsmath,amssymb,prl]{revtex4}
\usepackage{graphicx}
\usepackage{epsfig,amsmath,natbib}
\usepackage[hypertex]{hyperref}
\begin{document}
\title[]{Quantum capacitive phase detector}%Capacitive single Cooper-pair phase detector
\author{Leif Roschier }
\email{Leif.Roschier@iki.fi}
\author{Mika Sillanp\"{a}\"{a}}
\author{Pertti Hakonen}
\affiliation{Helsinki University of Technology, Low Temperature
Laboratory, P.O.BOX 2200, FIN-02015 HUT, Finland} 
\begin{abstract}
{We discuss how a single Cooper-pair transistor may be used to
detect the superconducting phase difference by using the phase
dependence of the input capacitance from gate to ground. The
proposed device has a low power dissipation because its operation
is in principle free from quasiparticle generation. According to
the sensitivity estimates, the device may be used for efficient
qubit readout in a galvanically isolated and symmetrized circuit.}
\end{abstract}

\maketitle

%\section{Introduction}
%RF-SET, LSET, CSEPD. All from SSET.
%The interest in Coulomb blockade and the accompanying quantum
%effects due to the superconductivity has triggered a wave of
%research.
The interest in Coulomb blockade and accompanying quantum effects
due to superconductivity has triggered a wave of research on new
physics and future applications. One basic device~\cite{averin86}
is the single-electron transistor (SET) which consists of two
small tunnel junctions having a sum capacitance of $C_{\Sigma}$.
If the single-electron charging energy $E_C = e^2 / (2
C_{\Sigma})$ dominates over temperature, $E_C \gg k_B T$, the SET
works as the most sensitive known electrometer.

In order to gain advantage of the inherently large bandwidth
$(R_{SET} C_{\Sigma})^{-1} \sim$ 10 GHz of the SET charge
detector, two new technologies have been developed where the SET
is read using an $LC$ oscillator. The RF-SET (radio-frequency SET)
is based on the gate dependence of the differential resistance of
a sequential tunneling SET, a property which modulates the
$Q$-value of the oscillator~\cite{schoelkopf}. No other device
than the RF-SET has been able to track dynamic single-charge
transport at MHz frequencies, which is relevant especially from
the point of view of characterization and eventual single-shot
readout of superconducting qubits~\cite{aassime}.

Because of the limitations due to the dissipative nature of the
RF-SET, a technique called L-SET (Inductive SET), having low
dissipation, has been developed recently~\cite{sillanpaa04}. With
zero DC-voltage, a superconducting SET, henceforth called single
Cooper-pair transistor (SCPT), behaves as an energy-storing
reactive component because of the Josephson coupling ($E_J / 2$ is
defined as the single-junction Josephson energy).

Since the first energy band $E_0$ of the SCPT grows approximately
quadratically as a function of the drain-source phase difference
$\phi$, the SCPT behaves as an inductor when looked at from the
source or drain. The effective Josephson inductance of a SCPT,
$L_{J} ^{-1} = \left( 2 \pi / \Phi_0 \right) ^2 \partial^2 E_0  /
\partial \phi^2$, has a strong dependence on the (reduced) gate
charge $n_g = C_g V_g/(2e)$ if $E_J / E_C \ll 1$. Here, $\Phi_0 =
h/(2 e)$ is the flux quantum. In the L-SET schematics, a charge
detector is built so that the resonance frequency of a system of a
SCPT and an $LC$ tank depends on $n_g$. So far, we consider the
L-SET the most promising method of sensitive and fast
electrometry.

\begin{figure}[htbp]
\begin{center}
\includegraphics[width=7cm]{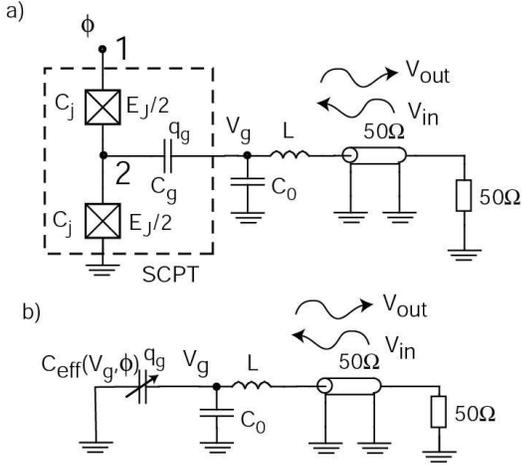}
\end{center}
\caption{a) CSET circuit. b) An equivalent circuit.}
\label{CSEPDskema}
\end{figure}

In addition to resistance or inductance, capacitance remains in
the group of linear circuit elements. The first energy band of a
SCPT grows approximately quadratically also with respect to the
second external parameter $n_g$. (A corresponding statement holds
also for a single junction, which has been suggested in Ref.\
\cite{averbrud} as a capacitance tunable by injected charge.)
Therefore, while observed from the gate electrode, the SCPT looks
like an effective capacitance $C_\mathrm{eff}$ to ground. The
operational principle of our proposal is based on the dependence
of $C_\mathrm{eff}$ on the source-drain phase difference $\phi$,

%As we will argue later in more detail, $C_\mathrm{eff}$ has strong dependence
%on the source-drain phase difference $\phi$, if $E_J / E_C \gg 1$.

\begin{equation}
C_\mathrm{eff}(\phi)^{-1} \equiv \frac{\partial V_g}{\partial
q_g(\phi)},
\end{equation}

\noindent where $q_g$ is the charge on the gate capacitor, as
illustrated in Fig.~\ref{CSEPDskema} (a), (b). Note that in
general $q_g \neq C_g V_g$.

Once coupled to a tank circuit, modulation of $C_\mathrm{eff}
(\phi)$ can be used to read a phase difference (henceforth, called
simply phase) reactively in a reflection measurement (see
Fig.~\ref{CSEPDskema}). Phase is defined as the time integral of
voltage, $\phi = (2 \pi / \Phi_0) \int_0^t V dt$. In the
superconducting case, $\phi$ equals the order parameter phase.

Unlike any previous considerations of single-electron or single
Cooper-pair devices, implementation of the "quantum capacitive
phase detector" device proposed in the present paper is a
generally fast and sensitive \emph{phase} detector. We call the
device a CSET, to emphasize its somewhat dual operation with
respect to the L-SET device.

%An inherent advantage of the proposed schematics is its symmetry
%with respect to the measured quantity, thus minimizing common-mode
%back-action due to the drive.

Hamiltonian for a SCPT symmetric in its Josephson energies is
~\cite{likharev85}
\begin{equation}\label{hamilt}
    H=E_{CP}\Big(\frac{\partial}{i\partial {\theta}}-n_g\Big)^2-E_J \cos ({\phi}/2) \cos
    ({\theta}),
\end{equation}

\noindent where a term $C_gV_g^2/2$ and terms having $\dot{\phi}$
have been ignored. $E_{CP}\equiv (2e)^2/(2C_\Sigma)$ is the
Cooper-pair charging energy and $E_J/2$ is the Josephson coupling
energy of the individual junctions. The phases are defined with
the help of voltages $V_i$ at points 1 and 2 in
Fig.~\ref{CSEPDskema}: $\theta=(2 \pi / \Phi_0) \int_0^t
(V_1/2-V_2) dt$ and $\phi=(2 \pi / \Phi_0) \int_0^t V_1 dt$. Here,
the former is the difference and the latter is the sum of the
phases over the two Josephson junctions.

\begin{figure}[htbp]
\begin{center}
\includegraphics[width=8cm]{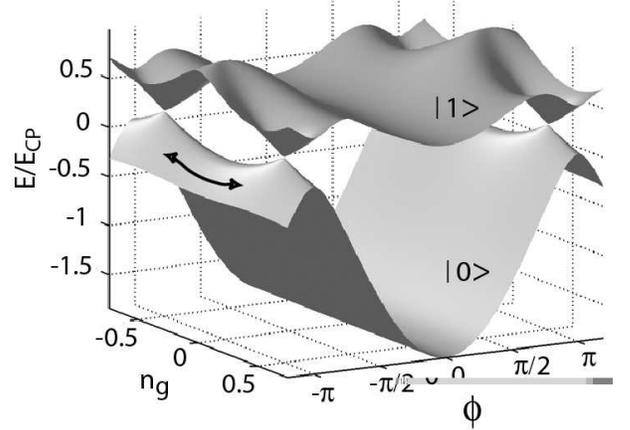}
\end{center}
\caption{The lowest ($|0\rangle$) and the first excited
($|1\rangle$) energy surface for a symmetric SCPT ($d = 0$) with
$E_J/E_{CP}$=3. The black line denotes the CSET operation point
${\phi}\sim \pi$ and $V_g(t)=V_{g0} \cos {\omega_0 t}$.}
\label{ejec3}
\end{figure}

The eigenvalues for this Hamiltonian are given by the solutions to
the Mathieu equation when $n_g \neq n/2$, where $n$ is integer
~\cite{likharev85,cottetthesis}:

\begin{widetext}
\begin{equation}\label{mathieueq}
    E_k({\phi},n_g)=\frac{E_{CP}}{4} \mathcal{M_A} \left(k+1-(k+1)[\mathrm{mod}\, 2]+2 n_g (-1)^k,
    \frac{2E_J \cos({\phi}/2)}{E_{CP}} \right),
\end{equation}
\end{widetext}

\noindent where $\mathcal{M_A}(r,q)$ is the characteristic value
$\mathcal{A}$ for even Mathieu functions~\cite{mathematica} with
characteristic exponent $r$ and parameter $q$. Unavoidable
asymmetry of tunnel junctions in a real device is easily
incorporated into Eq.\ \ref{mathieueq}. By substituting $E_J
\cos(\phi/2)$ in Eq.\ \ref{mathieueq} by $E_J
\sqrt{(1+d^2+(1-d^2)\cos(\phi))/2}$, we get energies of a SCPT
whose individual junctions have unequal Josephson energies
$E_J(1+d)/2$ and $E_J(1-d)/2$, where $d \neq 0$ is the asymmetry
parameter. Junction capacitances, however, can be arbitrarily
distributed. Fig.~\ref{ejec3} illustrates the two lowest
eigenvalues $E_0$ and $E_1$ with respect to the control parameters
${\phi}$ and $n_g$.

In order to calculate the observable capacitance $C_\mathrm{eff}$
from gate to ground when the system is in the lowest eigenstate
$E_0$, we first calculate $q_g = C_g (V_g - V_2)$, where the
island voltage is $V_2 = 1/C_g (\partial E_0 / \partial V_g)$.
Using Eq.~(1) we have

\begin{equation}\label{cccc}
    C_\mathrm{eff}=\frac{\partial}{\partial V_g}
    \left(C_g V_g - \frac{\partial E_0}{\partial V_g}\right)
= C_g - \frac{C_g^2}{C_Q},
\end{equation}

\noindent where $C_Q$ is the quantum capacitance $C_Q^{-1} \equiv
(\partial^2 E_0)/((2e)^2\partial n_g^2)$ due to the SCPT band
structure. In the following analysis, the constant term $C_g$ is
neglected, because it is small compared with the shunting
capacitance $C_0$.

\begin{figure}[htbp]
\begin{center}
\includegraphics[width=7cm]{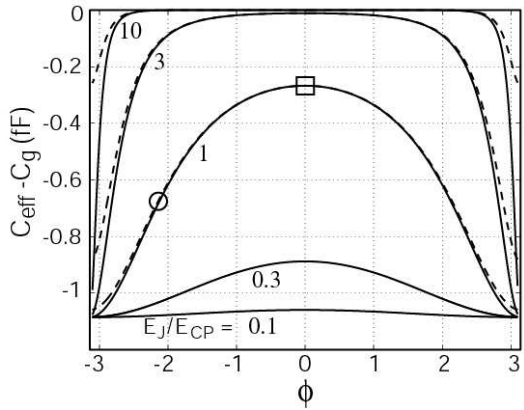}
\end{center}
\caption{Capacitances $C_\mathrm{eff}$ calculated using
Eq.(\ref{cccc}) with $C_g$ = 2 fF and $E_{CP}$=1 K. Solid lines
are for a symmetric SCPT ($d=0$), and the dashed lines for a
slightly asymmetric SCPT ($d = 0.1$). The numbers denote the
$E_J/E_{CP}$ ratio. The circle and the square correspond to the
operation points marked in Fig.~\ref{qubitvaihe} for the
charge-phase qubit readout.} \label{capesimerkki}
\end{figure}

In order to get maximum performance of the phase detector, we take
the operation point of the device such that the transfer function
$\partial C_\mathrm{eff}/\partial \phi$, which is the derivative
of the capacitance modulation curves in Fig.\ \ref{capesimerkki},
is maximized. This happens at ${\phi}$ rather close to $\pi$. As
seen in Fig.\ \ref{capesimerkki}, the transfer function increases
rapidly at large $E_J / E_{CP}$. The price to pay for a high gain
then is a limited dynamic range. The external gate drive is taken
as $V_g(t)=V_{g0} \cos {\omega_0 t}$.

Asymmetry in the SCPT weakens the modulation considerably at high
$E_J / E_{CP} \gtrsim 3$ as seen in Fig.\ \ref{capesimerkki}
(dashed lines). This is because asymmetry removes the degeneracy
at $n_g = \pm 0.5$ and $\phi = \pm \pi$ and smooths out the
strongest modulation.

As seen in Fig.\ \ref{CSEPDskema} (b), the proposed CSET circuit
is a series resonator with inductance $L$ and total capacitance $C
= C_0 + C_\mathrm{eff}$. Resonance frequency $f$ of the
configuration is then tunable by $C_\mathrm{eff}$. A shift of $f$
due to a change of capacitance $\Delta C$ is

\begin{equation}
f=f_0+\Delta f\propto\frac{1}{\sqrt{C_0+\Delta C}}\simeq
\frac{1}{\sqrt{C_0}}-\frac{\Delta C}{2C_0^{3/2}},
\end{equation}

\noindent and we find $\Delta f/f_0\simeq - \Delta C/2 C_0$.
$\Delta C$ can be written in terms of a phase change $\Delta \phi$
as $\Delta C = (\partial C_\mathrm{eff}/\partial \phi) \Delta
\phi$. For simplicity, we choose $\Delta \phi = 1$ radian.

We assume that the resonator itself does not dissipate power so
that its internal quality factor $Q_i = \infty$. The loaded
$Q$-factor $Q^{-1} = 1/Q_i + 1/Q_e$ is then set by the external
impedance $Z_0$ = 50 $\Omega$, so that $Q = Q_e = \omega L /Z_0 =
\sqrt{L/C}/Z_0$. This is a good approximation with low $Q$. It
follows that the phase $\vartheta=\arg(V_{out}/V_{in})$ of the
voltage wave amplitude reflected from the resonator, as
illustrated in Fig.~\ref{CSEPDskema}, changes by $\pi$ in the
frequency range of $\sim f_0/Q$. At the resonant frequency,
differential change of phase is larger, and by direct calculation
one gets the maximum modulation $\Delta \vartheta/\Delta f = 4
Q/f_0$, and

\begin{equation}
\Delta \vartheta \simeq - \frac{2 Q \Delta C}{C_0} = - \frac{2
\Delta C \sqrt{L}}{C_0^{3/2} Z_0} \label{dtheta}.
\end{equation}

\noindent We assume operation safely in the linear regime, such
that the gate charge amplitude is at maximum of the order $e/2$,
and thus the gate voltage amplitude is $V_{g0} \simeq e/2C_g$.
%\begin{equation}
%V_{g0} \simeq \frac{e}{2C_g}.
%\end{equation}
The average energy stored in the capacitors $C_\mathrm{eff}$ and
$C_0$ is of the order $E \simeq V_{g0}^2(C_\mathrm{eff}+C_0)/2 =
e^2 C /(8C_g^2)$.
%\begin{equation}
%E \simeq \frac{V_{g0}^2}{2} (C_\mathrm{eff}+C_0) =
%\frac{e^2}{8C_g^2}(C_\mathrm{eff}+C_0).
%\end{equation}
Because it takes $Q$ cycles for the (loaded) resonator to
dissipate most of its stored energy, the reflected power flow is
\begin{equation}
P \simeq E f_0/Q=\frac{f_0 e^2 C}{8Q C_g^2} \label{power}.
\end{equation}
This is about 5 fW with typical parameters in an experiment. The voltage
amplitude of the modulation is $\Delta V=\Delta \vartheta V_c$,
%\begin{equation}
%\Delta V=\Delta \vartheta V_c,
%\end{equation}
where the carrier voltage amplitude is $V_c=\sqrt{2 Z_0 P}$. The
modulation is conveniently transformed into power units to enable
later comparison with noise power. This leads to the definition of
information power:

\begin{equation}
P_{i}\equiv \frac{(\Delta V)^2}{2Z_0}=\frac{(\Delta \vartheta)^2
V_c^2}{2 Z_0}=(\Delta\vartheta)^2 P. \label{pi}
\end{equation}

\noindent By combining Eqs. (\ref{dtheta}), (\ref{power}) and
(\ref{pi}), we find

\begin{equation}
P_{i} \simeq Q \left(\frac{\Delta C}{C_0} \right)^2 \frac{f_0 e^2
C}{2 C_g^2}. \label{result}
\end{equation}

\noindent Eq.~(\ref{result}) may be written in terms of the
transfer function $\partial C_\mathrm{eff}/\partial \phi = C_g^2
\partial^3 E_0 / (4 e^2 \partial n_g^2 \partial \phi)$, $C_0$, and
$Z_0$. We also approximate $C_\mathrm{eff} \ll C_0$. It follows

\begin{equation}
P_{i}(\Delta \phi) \simeq \frac{1}{64 \pi Z_0 e^2}
\left(\frac{\partial^3 E_0}{\partial n_g^2
\partial \phi} \Delta \phi \right)^2 \left(\frac{C_g}{C_0}\right)^2. \label{result1}
\end{equation}

\begin{figure}[htbp]
\begin{center}
\includegraphics[width=8cm]{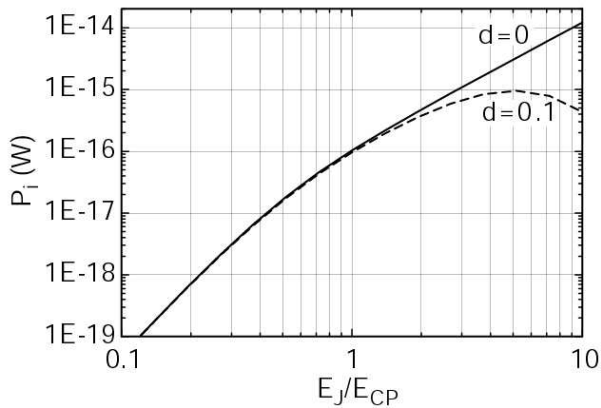}
\end{center}
 \caption{Information power (Eq.\ \ref{result1}) corresponding to a phase modulation
 $\Delta \phi = 1$ rad.
 Solid line: symmetric SCPT ($d=0$); dashed line: slightly
asymmetric SCPT ($d = 0.1$). Other parameters are as given in the
text.} \label{kuvaajat}
\end{figure}

\noindent Sensitivity increases dramatically by increasing the
$E_J/E_{CP}$ ratio as is illustrated in Fig.~\ref{kuvaajat}.
$E_{CP} \gg k_B T$ must be satisfied in order to keep the system
localized in the lowest state $E_0$. At a fixed ratio of
$E_J/E_{CP}$, maximization of $C_g / C_0$, and minimization of
$Z_0$ yields the best sensitivity.

In order to estimate device performance we take the operation
frequency $f_0$ to be $\sim$~1 GHz. The individual component
values are chosen to be easily realizable by present fabrication
technology. Due to stray capacitance from bonding pads etc., it is
difficult to achieve in practice $C_0$ lower than $\sim 0.15$ pF.
To have a series resonance at the $f_0$ with $Q$-factor of
$\sim$20, this would be accompanied by an inductance $L \sim 160$
nH. The inductance may be easily realized by using a commercial
surface mount coil. It is to be noted that the capacitance
$C_\mathrm{eff}$ is $\sim$ 1/75 of the capacitance $C_0$ assuming
a large $C_g \sim$ 2 fF. The large gate capacitance is easily
implemented with a similar overlap junction as the two SCPT tunnel
junctions, but using a much longer oxidation to create a highly
resistive junction~\cite{zimmerli92}.

For phase sensitivity estimates, we calculated $\Delta C/C_0$
resulting from a phase change $\Delta \phi = 1$ rad, at different
ratios of $E_J/E_{CP}$, and assuming $E_{CP} = 1$ K. They are
listed in Table~\ref{tab1}. The values are calculated at the
optimal operation point of $\phi$ which maximizes the transfer
function, and at $V_g \sim 0$. As seen in Fig.~\ref{kuvaajat}, the
information power is $P_i \sim 10$ fW with $E_J/E_{CP} \sim 10$.

By using a cryogenic HEMT amplifier with a typical noise
temperature of 3 K that corresponds to spectral density of $s_N
\sim 4\times$10$^{-23}$ W/Hz, we find the phase sensitivity
$s_\phi = \sqrt{s_N / P_i(1 \rm rad)}$ of the order 45 $\mu{\rm
rad}/\sqrt{\rm Hz}$ for symmetric SCPT. Combined with a low-noise
SQUID amplifier~\cite{muck} that would be close to the resonator
circuit, the value $Z_0$ could be lowered to a value of 1 $\Omega$
or below~\cite{Zin} and the noise temperature could be down by a
factor 30. This would result in a sensitivity of 1.3 $\mu{\rm
rad}/\sqrt{\rm Hz}$.

Asymmetry in the SCPT junctions weakens the numbers at high
$E_J/E_{CP}$, as portrayed by the dashed line in Fig.\
\ref{kuvaajat}. With $E_J/E_{CP} = 10$ and $d = 0.1$, the
mentioned sensitivities would go down by a factor of 5. However,
at high $E_J/E_{CP}$ the junctions are naturally of large area,
and hence relatively easy to fabricate with similar sizes.

\begin{table}
\caption{Linearized relative change of capacitance corresponding
to a $\Delta \phi = 1$ rad, for symmetric SCPT (parameters as in
the text).} \label{tab1}
\begin{center}
\begin{tabular}{c|ccccc}
  \hline \hline
  $E_J/E_{CP}$                                 & 0.1   & 0.3 &1    & 3     & 10  \\
  $\Delta C/C_0$    & $8.5$e$-5$ &$6.6$e$-4$ &$3.6$e$-3$ &$1.1$e$-2$ &$4.3$e$-2$
  \\
  \hline \hline
\end{tabular}
\end{center}
\end{table}

For the estimates discussed here, we have assumed operation of the
device only at "safe" values of $V_g$. That is, variation of the
capacitance $C_\mathrm{eff}$ is continuous over a reasonable range
of values around the operation point. As evident in
Fig.~\ref{ejec3}, close to the degeneracy points, the differential
change of capacitance is substantially larger. One needs, however,
a relatively strong AC gate drive $V_{g0}$ in order to distinguish
the carrier signal from the pre-amplifier noise power within the
bandwidth. For this reason, we have omitted the analysis of these
points.
%\section{Readout of charge-phase qubit}

\begin{figure}[htbp]
\begin{center}
\includegraphics[width=8cm]{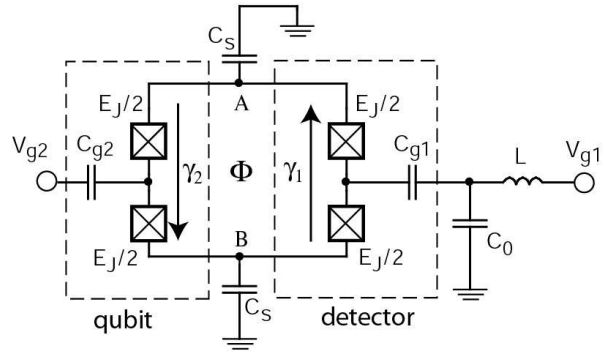}
\end{center}
\caption{Schematic of the CSET coupled to a charge-phase
qubit~\cite{vion02}. The gauge-invariant phase differences
$\gamma_i$ and their orientations are marked with arrows. All
Josephson junctions are taken to have same Josephson coupling
energy $E_J/2$. $\Phi$ is the externally applied magnetic flux.}
\label{qubitluku}
\end{figure}

In order to demonstrate an important application of the CSET, we
discuss a circuit, where the CSET works as a detector for the state
of the charge-phase qubit (see Fig.~\ref{qubitluku}).

The basic idea of this circuit is that since the qubit states
$|0\rangle$ and $|1\rangle$ correspond to distinct phases $\phi$,
they result in a different capacitance of the CSET phase detector.

Due to a supposed large capacitance, $C_s \gtrsim 20$ fF, phases
with respect to ground at points A and B are taken as classical
variables. Charging energy of these leads is neglected for the
same reason. The structure having this high capacitance is
naturally fabricated on a ground plane substrate. For a standard
insulator thickness of 300 nm on top of a conducting ground plane,
the loop structure would measure only tens of $\mu$m in size.

We assume that the circuit can be described as two connected SCPTs
(detector and qubit) and that the gauge-invariant phase $\gamma_i$
over each of them is found by solving the equations
\begin{equation}
\gamma_1+\gamma_2=-2\pi \Phi/\Phi_0(\mathrm{mod}\, 2\pi),
\label{class1}
\end{equation}
 and
\begin{equation}
\frac{\partial E_{k'1}(\gamma_1,V_{g1})}{\partial
\gamma_1}=\frac{\partial E_{k2}(\gamma_2,V_{g2})}{\partial
\gamma_2}.\label{class2}
\end{equation}
The former equation assures the correct $2\pi$-periodicity of the
sum of the phases around the loop. The latter equation states that
the currents flowing through both SCPTs have the same magnitude.
The energies $E$ are defined according to Eq.~(\ref{mathieueq}).
The band index $k'$ for the detector is always null and the qubit
index $k$ takes values 0 or 1, corresponding to the states
$|0\rangle$ and $|1\rangle$. For this example we assume for
simplicity that both SCPTs have $E_J/E_{CP}=1$.

As an example of the operation we suppose that after manipulating
the qubit, it is in a superposition which then collapses into
either $|0\rangle$ or $|1\rangle$ when the detector is turned on.
The detector, on the other hand, stays in the state $|0\rangle$.
In order to start the measurement, the qubit gate voltage is
turned to approximately $C_g V_{g2}/2e$ = 0.37. The detector is
maintained at $C_g V_{g1}/2e$ = 0 throughout the operations.
Thereafter, an externally applied magnetic flux is ramped
adiabatically to $\Phi/\Phi_0 \approx 1/2$.

\begin{figure}[htbp]
\begin{center}
\includegraphics[width=8cm]{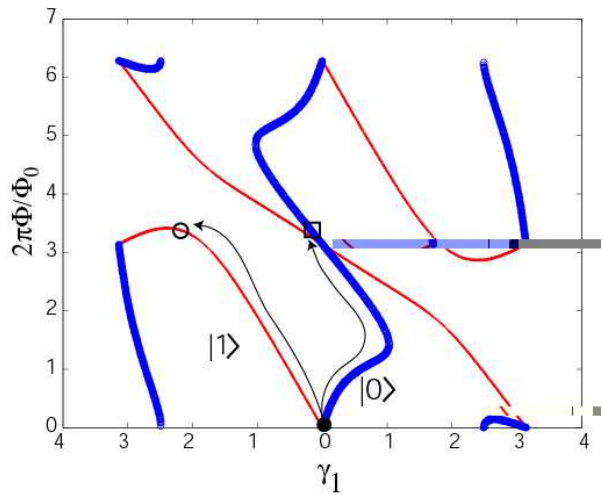}
\end{center}
 \caption{Dependence of the phase over the detector $\gamma_1$
on the applied magnetic flux $\Phi$. The thick (blue) lines
correspond to the solutions of Eqs.~(\ref{class1}) and
(\ref{class2}), when the qubit is in the state $|0\rangle$. The
thin (red) lines are for the qubit state $|1\rangle$,
respectively.} \label{qubitvaihe}
\end{figure}

Fig.~\ref{qubitvaihe} illustrates the dependence of $\gamma_1$
with respect to the qubit state. Depending on the state of the
qubit, the phase over the detector circuit $\gamma_1$ will thus
become either 0 or $\sim$ $-2$ rad. According to
Fig.~\ref{capesimerkki}, the capacitance $C_\mathrm{eff}$ then
differs by $\sim 0.4$ fF, depending on the qubit final state.
According to Eq.~(\ref{result}), the difference corresponds to an
information power of $-132$ dBm, by using easily achievable values
$C_0$=0.15 pF, $Q$=20, $f_0$=1 GHz and $C_g$=2 fF.
Signal-to-noise-ratio 1 is then achieved at a bandwidth of 20 MHz
with a feasible SQUID first stage amplifier having a noise
temperature of 0.3 K. Measurement is thus clearly possible in
sub-microsecond regime. A reasonable junction asymmetry, $d = 0.1$
has a negligible influence on the curves in Fig.\
\ref{qubitvaihe}.

The proposed qubit circuitry has several important advantages.
%Some of them are apparently not addressed in previous qubit
%proposals or realizations.
First, low power dissipation means low rate of quasiparticle
generation, which is essential for low back-action and fast
recovery from the measurement. Second, since the circuit is
galvanically isolated, it is free from external quasiparticle
injection.

The third advantage is the symmetry of the schematics. Although in
some sense the circuit is strongly coupled to the external $Z_0$
via the detector gate, thermal noise of $Z_0$ acts only as a
common-mode signal. This is equivalent to say that the real part
of the impedance seen by the qubit $\rm Re (Z)$ is very small.
Asymmetry weakens the situation, but with a realistic, random
asymmetry of about 20 \% in the component values, we calculated
the following figures in the schematics of Fig.\ \ref{qubitluku}:
$\rm Re (Z) \ll 1$ m$\Omega$ both at low frequency ($f < 1$ GHz)
which is relevant for dephasing, and at the level-spacing
frequency ($10-50$ GHz) which affects relaxation. Especially since
the qubit operations are naturally performed at the saddle point
$\phi = 0$, $n_g = \pm 1/2$ similarly as in the original
charge-phase qubit, the system is extremely well decoupled from
the environment. Note that the gate and flux operation leads of
the qubit may be weakly coupled so that they do not contribute
noise.

Coherence time is then presumably limited by internal $1/f$ noise,
which should be similar to existing qubit realizations. Its effect
weakens as $E_{CP}$ grows. At a conservative value $E_{CP} = 1$ K,
we estimate a dephasing time~\cite{cottet1perf} of $1-2 \; \mu$s,
which is comparable to the original charge-phase qubit.

In conclusion, we have proposed a technique to measure the
superconducting phase difference by monitoring the effective
capacitance between the gate of a single Cooper-pair transistor
and ground. As a practical example, the readout of a charge-phase
qubit using the technique was discussed.

Fruitful discussions with T. Heikkil\"{a}, N. Kopnin and T.
Lehtinen are gratefully acknowledged. This work was supported by
the Academy of Finland and by the Large Scale Installation Program
ULTI-3 of the European Union.

\bibliography{LEIFBIB}

\begin{thebibliography}{13}
\bibitem[{Zin()}]{Zin}
\bibinfo{note}{Input impedance of microstrip SQUID amplifier is determined by
  the shunt resistors transformed into amplifier input. Also, a non-standard
  way of coupling the input signal is required in the application.}
\end{thebibliography}

\end{document}